\documentstyle[preprint,prl,aps]{revtex}
   \newcommand{\mra}  {\rightarrow}
   \newcommand{\vecbm}[1]{\mbox{\boldmath#1}}
   \newcommand{\vecb}[1]{\mbox{\bf#1}}
   \newcommand{\triplint}{\int\rule{-3.5mm}{0mm}\int\rule{-3.5mm}{0mm}\int}
   \newcommand{\doublint}{\int\rule{-3.5mm}{0mm}\int}
   
   \newcommand{\mlora} {\longrightarrow}
\begin{document}
%%\draft
%\tighten
\title{
Fragmentation Phase Transition in Atomic Clusters I\\
--- Microcanonical Thermodynamics ---}
\author{
D.H.E. Gross, M.E. Madjet, and O. Schapiro}
\address{
Hahn-Meitner-Institut
Berlin, Bereich Theoretische Physik, \\14109 Berlin, Germany\\
and\\
Freie Universit\"at Berlin}
%\today}
\maketitle
\begin{abstract}
The volume $W$ of the accessible N-body phase space and its dependence on the
total energy is directly calculated. The famous Boltzmann relation $S=k*ln(W)$
defines microcanonical thermodynamics {\em (MT)}. We study how phase
transitions appear in {\em MT}. Here we first develop the thermodynamics of
microcanonical phase transitions of first and second order in systems which are
thermodynamically stable in the sense of van Hove.  We show how both kinds of
phase transitions can unambiguously be identified in relatively {\em small}
isolated systems of $\sim 100$ atoms by the shape of the microcanonical caloric
equation of state $<T(E/N)>$ and {\em not} so well by the coexistence of two
spatially clearly separated phases. I.e.{\em within microcanonical
thermodynamics one does not need to go to the thermodynamic limit in order to
identify phase transitions.} In contrast to ordinary (canonical) thermodynamics
of the bulk microcanonical thermodynamics {\em (MT)} gives an insight into the
coexistence region. Here the form of the specific heat $c(E/N)$ connects
transitions of first and second order in a natural way. The essential three
parameters which identify the transition to be of first order, the transition
temperature $T_{tr}$, the latent heat $q_{lat}$, and the interphase surface
entropy $\Delta s_{surf}$ can very well be determined in relatively small systems
like clusters by {\em MT}. It turns out to be essential whether the cluster is
studied canonically at constant temperature or microcanonically at constant
energy.  Especially the study of phase separations like solid and liquid or, as
studied here, liquid and gas is very natural in the microcanonical ensemble,
whereas phase separations become exponentially suppressed within the canonical
description.  The phase transition towards fragmentation is introduced. The
general features of {\em MT} as applied to the fragmentation of atomic clusters
are discussed.  The similarities and differences to the boiling of macrosystems
are pointed out.
\end{abstract}
\section{Introduction} 
This is the first paper in a series of papers treating the topological
structure of the N-body phase space of atomic clusters by microcanonical
thermodynamics ({\em MT}). It was one of the primary issues of cluster physics
to understand macrosystems starting from the properties and interactions of its
constituents, the single atoms or molecules.  Especially one likes to
understand how the most spectacular changes in the thermodynamical behavior of
macro systems, phase transitions, develop with increasing number of atoms from
atomic dimensions towards the bulk. However, small clusters have also their
specific features. They can be charged, their shape (surface) is important,
they can fragment, they can rotate, etc.

Studying clusters as thermodynamical systems we consider the phase space which
can be reached by all constituents of the cluster. The topology of this total
accessible N-body phase space reflects (or implies?) the behavior of many
interacting systems because the dynamics is often chaotic. Then the evolution
of many replica of the same system under identical macroscopic initial
conditions follows the structure of the underlying N-body phase space. It is
ergodic. In nuclear fragmentation the ergodicity is presumably due to the
strong and short ranged friction between moving nuclei in close proximity.
Friction between atomic clusters is yet unknown but probably it exists there
also. 

First we have to discuss the concept of thermodynamics of small systems in
general and especially of phase transitions. 
It is important to realize that isolated clusters must be treated {\em
microcanonically.} Usually there is no external heat- or particle bath which
defines temperature, pressure, or chemical potential\footnote{It is possible to
prepare atomic clusters initially in a canonical ensemble by embedding them in
an inert carrier gas \protect\cite{ellert95}. However, most probe-reactions on
these clusters are too fast to keep the cluster during this time at constant
temperature.  Moreover frequent collisions with the carrier gas will distort
the cluster so that one would not measure the properties of the cluster
alone.}.  Microcanonical thermodynamics is the proper theory for isolated small
systems.

Phase transitions are well defined as singularities in various canonical or
grandcanonical expectation values  as function of e.g. the temperature in the
thermodynamic limit $\lim_{N \mra\infty}|_{N/V=\varrho}$. E.g. the melting
transition of bulk sodium at normal pressure shows up as a peak in the specific
heat in the neighborhood of $T_m= 371$K on top of which there is a
$q_{lat}\delta(T-T_m)$ singularity c.f. fig.\ref{cvNa}. With the normal finite
resolution the delta-function manifests itself by a jump in the bulk-specific
heat by $q_{lat}$, the specific latent heat, see fig.\ref{cvNa}. There are
different ways to identify the phase transiton of melting: First one realises
that the systems changes its stucture, it melts. Another way is to measure
e.g. the specific heat $c(T)$ or the caloric equation of state $<T(E/N)>$ and
realizing that at the melting the temperature does not rise if more energy is
put into the system up to the point where the whole system becomes liquid. I.e.
that the caloric equation of state $<T(E/N)>$ has an anomaly (here a plateau).

The basic questions are now: Can one define phase transitions in finite,
eventually small systems like atomic clusters. Is it possible to define phase
transitions of small isolated systems microcanonically as function of the
energy not as function of the temperature?  From which size on do phase
transitions exist in such small systems? Is a cluster of some 50 to 100 atoms
big enough?  We will show that it is in fact much easier to identify phase
transitions in the microcanonical ensemble and that this is possible in
astonishingly small systems. Small clusters can undergo a phase transition and
one can unambiguously distinguish continuous (second order) from discontinuous
(first order) transitions by the form of the microcanonical caloric equation of
state $<T(E/N)>$. We will see that the main signal of a transition in finite
systems is {\em not} the singularity of some canonical expectation value in the
thermodynamic limit but a specific anomaly in the smooth microcanonical
equation of state $<T(E/N)>$ which can be well identified already in pretty small
systems. Of course, the transition may change its character when the number of
particles increases. However, in the case of the 10-states Potts model as well
in the case of the boiling of sodium or potassium the bulk values of the three
quantities above are clearly approached by the corresponding transitions in
small systems/clusters \cite{gross157} of some $100$ atoms.

A microcanonical ensemble has some peculiarities: It does not have a positive
definite heat capacity. In fact at a phase transition of first order the
specific heat $c(\varepsilon)$ becomes {\em negative} in general.  Therefore
the classical signal of a peak in the specific heat is not useful to
characterize a phase transition in small systems.

The second peculiarity of small bound systems like atomic clusters is a new
structural phase transition which does not exist in infinite homogeneous
systems: At higher excitation they often do not simply boil into a gas of
monomers. On the contrary, they may {\em fragment} into few relatively large
fragments.  In this case the size of the droplets at the transition point can
be of the order of the size of the system itself e.g. in the case of fission.
The fluctuation of the fragment sizes is substantial. Then one cannot ignore
the droplets compared to the monatomic vapor anymore. The phase transition is
not determined alone by the equilibrium of the homogeneous liquid with the
homogeneous monatomic gas controlled by the equality of the chemical potentials
of liquid and gas as we are used to in conventional (grandcanonical)
thermodynamics.  Often the number of droplets is similar to the number of
monomers or even larger. Such a process was found recently in the decay of hot
nuclei and is called multifragmentation.

It is one of the aims of this series of papers to study the conditions for
fragmentation of hot and/or highly charged atomic clusters.  However, also
rapidly rotating clusters could fragment or even fission symmetrically.
Therefore it is important to develop microcanonical thermodynamics including
angular momentum as a fundamentally conserved quantity or order parameter.

As today one is not familiar to microcanonical thermodynamics we will discuss
in the following sections some of its basic features and show how phase
coexistence and phase separations show up in {\em MT}. Canonical and
grandcanonical ensembles have the great advantage that one can treat many
simple examples like systems which can be transformed into ideal noninteracting
gases analytical. The prize one has to pay is that both ensembles average in
the limit of large particle numbers over spatial inhomogeneities like e.g.
phase separation and interphase surfaces. Only under simple geometrical
conditions this defect can be cured by the Maxwell construction. In contrast to
this {\em MT} allows for such inhomogenuous situtions in a natural way.
\section{Microcanonical thermodynamics}
Statistical mechanics as foundation of thermodynamics goes back to the
works of Boltzmann and of Gibbs. Boltzmann's gravestone has the famous
epigraph:
\begin{center}\fbox{\fbox{\vecbm{$S=k*lnW$}}}\end{center}
which is the most concise formulation of the essence of thermodynamics.
$W$ is the volume of the N-body phase space of the system (here in units of
($(2\pi
\hbar)^{3N}$) or the partition sum of all possible quantum states of the
N-body system:
\begin{equation} 
\Omega_N(E)=\sum_{\nu}{\delta_{E,E_{\nu}}} \label{partsum},
\end{equation}
and $S$ is the N-body entropy of the system.
 
Microcanonical thermodynamics explores the topology of the N-body phase space
and determines how $\Omega_N$ depends on the fundamental globally conserved
quantities namely total energy $E=N*\varepsilon$, angular  momentum $\vecb{L}$,
mass (number of atoms) $N$, charge $Z$, linear momentum $\vecb{p}$, and last
not least the available spatial volume $V$ of the system. This definition is
the basic starting point of any thermodynamics since Boltzmann, see e.g.
\cite{thompson88}. In his
``Vorlesungen \"uber Gastheorie, II. Teil'' he calls the microcanonical
ensemble the ``Ergode''\cite{boltzmann}. 
It is an entirely mechanistic
definition saying that if we do not know anything more about a complicated
interacting N-body system but the values of its globally conserved macroscopic
parameters the probability to find it in a special phase space point (N-body
quantum state) is uniform in the accessible phase space.

The entropy is defined as the logarithm of $\Omega$ see Boltzmann's epigraph
\begin{equation}
S(E,V,N)=Ns(\varepsilon=E/N)=ln(\Omega(E,V,N)),\label{entropy}
\end{equation} we take Boltzmann's constant $k=1$. 
The thermodynamic temperature $T_{thd}$ is defined by
\begin{eqnarray}
\beta&=&\frac{\partial S(E,V,N)}{\partial E}= \frac{\partial s(\varepsilon)}
{\partial \varepsilon} \label{beta} ,\\
T_{thd}&=&\frac{1}{\beta} .\label{temperature}
\end{eqnarray}
By Laplace transform of $\Omega(E,V,N)$ one steps from the ``extensive''
variables like 
$E,V,N$ to the intensive ones like $T,P,\mu$. E.g. the Gibbs grand partition
function and the Gibbs grand potential are then
\begin{eqnarray} 
Z(\beta,P,\mu)&=&
\triplint_0^{\infty}{\Omega(E,V,N)e^{-\beta(E+PV-\mu N)}\;dE\;dV\;dN} ,\\ 
G(\beta,P,\mu)&=&-T\; ln[Z(\beta,P,\mu)].
\end{eqnarray} I.e. Any sufficiently isolated part of the system has the {\em
same} grand partition function $G(T,P,\mu)$, i.e. the system becomes
{\em translationally invariant or homogeneous} in the bulk limit. In the same
way one gets the canonical partition function and the Gibbs free energy as
\begin{eqnarray}
Z(\beta,P,N)&=&\doublint_0^{\infty}{\Omega(E,V,N)e^{-\beta N(\varepsilon+Pv)}\;dE\;dV}
,\label{laplace}\\ 
\mbox{GF}(T,P,N)&=&-T\; ln[Z(\beta,P,N)]. 
\end{eqnarray}

Phase transitions in atomic clusters were quite early investigated
microcanonically by molecular dynamical simulation of e.g. the melting
transition in Ar$_{13}$ clusters see the pioneering work by Jellinek, Beck and
Berry \cite{jellinek86}. In these early calculations the transition energy
could be fixed by the anomaly of the ``caloric'' curve E$_{kin}/N$ vs.
E$_{tot}/N$. Neither the entropy, nor the proper thermodynamic temperature, nor
the order of the transition was determined. In contrast to these early work we
will here present a proper calculation of the volume $W$ resp.  $\Omega_N$ of
the accessible N-body phase space and its dependence on the total energy. We
thus follow literally the prescription of Boltzmann's epigraph. This way, we
will be able to determine all decisive parameters of the phase transition in
question: $T_{tr}$, $q_{lat}$, $\sigma_{surf}$ and $\Delta s_{boil}$, the gain
of entropy per evaporated atom, for the clusters.

\section{Differences between microcanonical and canonical ensemble}

According to van Hove  a system of $N$ particles interacting
via short range two-body  attractive forces with hard cores is
thermodynamically stable, the thermodynamic limit of $N, V
\mra\infty|_{N/V=\varrho}$ exists for such systems, intensive quantities like
the specific energy have finite limiting values \cite{vanhove49} see also
\cite{thompson88}. Then, the thermodynamics derived from the microcanonical
partition sum $\Omega(E,N,V)$ and the one derived from the canonical
$Z(\beta,P,N)$ or the grand canonical partition function $Z(\beta,P,\mu)$
(usually) coincide. Outside of phase transitions of first order, the relative
fluctuations $\Delta E/E$, or $\Delta\varrho/\varrho$ vanish $\propto
1/\sqrt{N}$.

This is naturally quite different for {\em finite} systems. However, even for
bulk systems {\em at} phase transitions ($T=T_{tr}$) of first order do the
microcanonical and the (grand)canonical ensemble differ essentially.  In the
(grand)canonical ensemble the energy fluctuations $\Delta \varepsilon$ per
particle remain finite even in the thermodynamic limit
($(\Delta\varepsilon)^2|_{T_{tr}}
\propto q_{lat}^2$, the specific latent heat).  Consequently, the difference
between the microcanonical and the (grand)canonical ensemble persists at
transitions of first order and we must expect both ensembles to describe
different physical situations. In section \ref{firstsecond} and especially in
the subsection \ref{phsep} we will illuminate this important point in more
detail.

Systems interacting via long range forces like unscreened Coulomb repulsion
between charged parts of equal sign, or the centrifugal force when they are
rotating, or under the long range gravity, don't have a thermodynamic limit and
must be described by the microcanonical ensemble. Such systems fragment
macroscopically into, in general, several regions of high density --- condensed
matter --- and also into, in general, several regions of low density --- vapor
or may be empty space. Even though in the first two examples, Coulomb repulsion
or repulsive centrifugal force, the systems are unstable they may
disintegrate initially slowly enough to establish a transient statistical
equilibrium. This is evidently the case in the fragmentation of hot nuclei and
we expect this to be quite similar in the case of fragmentation of atomic
clusters.  Differently from conventional thermodynamics where systems must be
in a homogeneous phase at fixed temperature everywhere, here the system is most
likely inhomogeneous.  The inhomogeneities and their fluctuations are more
important in characterizing the state of the system than any mean values.  {\em
In contrast to thermodynamics of the homogeneous bulk, in small systems or
large systems with long range forces the entropy connected to different
partitions of the system is an important part of the total entropy.} Familiar
formulas like the one-particle entropy
\begin{equation}
s_{sp}= -\sum_a{n_a ln(n_a)}
\end{equation}
are useless for calculating the total entropy.

\section{First and second order phase transitions in small systems}
\label{firstsecond} 
 
Macroscopic systems have a discontinuity in the specific heat $c_{bulk}(T)$ at
first order phase transitions. $c_{bulk}(T)$ may have a finite peak at
$T\approx T_{tr}$.  On top of this there is a spike $q_{lat}\delta(T-T_{tr})$.
With finite resolution it shows up as a jump in $c_{bulk}(T)$ by the latent
heat $q_{lat}$.  A typical example is the specific heat of bulk sodium at the
melting transition, fig.\ref{cvNa}.  
In contrast, a transition of second order is
continuous at the transition temperature where $c_{bulk}(T)$ has (in the
example of the Ising model) a logarithmic singularity in $T-T_{tr}$.

Microcanonical thermodynamics gives new and deep insight into this. It will
further allow to extend the concept of phase transitions to systems not
treated before by thermodynamics like systems with long range forces or
strongly rotating systems. We begin with the discussion of microcanonical phase
transitions in standard model systems in which phase transitions of first and
second order are well known. As an example we take the 2-dimensional 10-states
Potts model for which the asymptotic thermodynamics is even known analytically
\cite{baxter73}. The q-states Potts model is a generalization of the Ising
model by allowing $q$ instead of only $2$ spin values at each lattice point. In
two dimensions it has a second order phase transition for $q\le4$ and a first
order transition for $q>4$ from a spin-ordered (in the following often called
colloquially ``liquid'') to spin-disordered phase (called ``gas'').

In ref.\cite{gross150} we determined the three basic parameters of phase
transitions of first order within microcanonical thermodynamics, the transition
temperature $T_{tr}$, the specific latent heat $q_{lat}$, and the specific
interphase surface entropy $\Delta s_{surf}$ for the 10-states Potts model,
i.e. a systems with a nearest neighbor
coupling. It was demonstrated that for surprisingly small systems the values
of these three parameters are closer to their asymptotic values than in the
canonical ensemble. This is so because most of the finite-size scaling is due
to the large, but trivial, exponent in the Laplace transform,
eq.(\ref{laplace}), from the microcanonical to the canonical partition
sum \cite{hueller94,promberger96}.

From our studies of the Potts model \cite{gross150} we made the plausible
conjecture, which will serve us as working hypothesis: 
{\em The two types of phase transitions in thermodynamically stable finite
systems are distinguished by the form of the microcanonical caloric equation of
state $T_{thd}(\varepsilon)$: A transition of first order has a {\em
backbending} caloric equation of state $T^{-1}=\beta(\varepsilon)$, see figure
\ref{entrop}b. In contrast a phase transition of second order has only a
horizontal saddle point in $<T(E/N)>$.}
For a system with nearest neighbor interactions the area between
$\beta(\varepsilon)$ and the ``Maxwell'' line $\beta_{tr}=1/T_{tr}$ is twice
the interphase surface entropy $\Delta s_{surf}$ \cite{gross150} and the
coefficient of surface tension is $\sigma_{surf}=T_{tr}N^{1/3}\Delta s_{surf}$.
Following M.Fisher a nonvanishing surface tension qualifies the transition to
be of first order
\cite{fisher67}, in other words our experience with the Potts model suggests
that the backbending (or S-bend) \cite{doye95b,kunz94,wales95a} is a {\em 
necessary} signature for a phase transition of first order in a finite
thermodynamically stable system. It is interesting to notice that the
suppression of the configurations with phase coexistence compared to those with
pure phases in the canonical ensemble is used since long to calculate the
surface tension in general thermodynamical stable systems cf.e.g.
\cite{binder82}:
\begin{eqnarray}
P(\mbox{pure})&=&\frac{e^{-\mbox{\scriptsize GF}_{pure}/T}}{Z(T)}\\
P(\mbox{mix})&=&\frac{e^{-\mbox{\scriptsize GF}_{mix}/T}}{Z(T)}\\
\mbox{GF}_{mix}&=&\mbox{GF}_{pure}+\mbox{GF}_{surf}\\
\mbox{GF}_{surf}&=&N^{2/3}\sigma_{surf}\\
\frac{P_{tr}(\mbox{mix})}{\sqrt{P_{tr}(\mbox{phase1})P_{tr}(\mbox{phase2})}}&=&
e^{-N*\Delta s_{surf}}=e^{-N^{2/3}\sigma_{surf}/T_{tr}},
\label{binder}
\end{eqnarray}
where $P(\mbox{pure})$ is the probability to find the system at $T_{tr}$ in one
of the pure phases, whereas $P(\mbox{mix})$ is the probability to find it in
configurations with two separated phases. This generalizes to other systems our
conjecture that the backbending of $T(\varepsilon)$ is a neccessary condition
for the surface tension and for a transition of first order. Of course, there
may also be other reasons for a backbending of $T(\varepsilon)$.

The left darkened area is the defect of entropy $\Delta
s_{surf}=\int_{\varepsilon_1}^{\varepsilon_2} \beta d\varepsilon$ that the
system `pays' for introducing interphase surfaces, which it finally gets back
when the whole system is converted to the new phase at
$\varepsilon=\varepsilon_3$, right darkened area, and the interphase surface
disappears \cite{gross150}. In the bulk the transition is discontinuous as
function of $T$ or $\beta$ and ``jumps'' from the liquid
($\varepsilon\le\varepsilon_1$) to the gas branch
($\varepsilon\ge\varepsilon_3$) of the caloric curve, fig.\ref{entrop}. As a
function of the specific energy $\varepsilon$ the transition is however
continuous. With rising $\varepsilon$ the system passes smoothly from the
liquid phase over a mixed phase with large coexisting fluctuations of the two
phases (``gas bubbles'' and ``liquid droplets'') towards the pure gas phase
when the specific energy is increased by the specific latent heat $q_{lat}$.
Figure
\ref{entrop}a shows the corresponding specific entropy
$s(\varepsilon)=\int_0^{\varepsilon}{\beta(\varepsilon')d\varepsilon'}$.      
Notice that the entropy is a monotonously rising function with rising energy in
the region of energies shown. 
In figure \ref{entrop}a we subtracted a linear function $a+b\varepsilon$ in   
order to visualize the transition. Otherwise, on the scale                    
of fig.\ref{entrop}a one would not be able to distinguish $s(\varepsilon)$    
from a straight line.                                                         
The transition is characterized by the convex intruder in $s(\varepsilon)$ of
depth $\Delta s_{surf}$.  The effect of the interphase surface tension is a
somewhat slower increase of the entropy with rising specific energy as the
system prefers to create interfaces between regions of different phases,
spin-aligned and disordered ones.  Figure \ref{entrop}c shows the specific heat
capacity
\begin{equation}
c(\varepsilon)=\frac{\partial\varepsilon}{\partial<T>}=\frac{-\beta^2}
{\partial\beta/\partial\varepsilon}
\end{equation}
as a function of the specific energy $\varepsilon$. (Here numerical
fluctuations in $\beta(\varepsilon)$ in figure \ref{entrop}b have been
smoothed). One can see {\em within the coexistence region of
$\varepsilon_1\le\varepsilon\le\varepsilon_3$ (shaded area in
fig.\ref{entrop}), the microcanonical specific heat has two poles and becomes
even negative in between.}

The convex intruder in the specific entropy $s(\varepsilon)$,
fig.\ref{entrop}a, is forbidden by van Hove's theorem in the canonical ensemble
for an infinite number of particles \cite{vanhove49} see also e.g.
\cite{thompson88}. At energies in the region 
of the intruder, an infinite system is unstable against spontaneous devision
into two parts \cite{hueller94}. Conventional thermodynamics of the bulk is
blind in this energy interval. The Laplace transform eq.(\ref{laplace}) has no
stationary point on the branch with the positive slope of $\beta(\varepsilon)$.
It can only see the branches of $c(\varepsilon)$ in the regions
$\varepsilon\le\varepsilon_1$ and $\varepsilon\ge\varepsilon_3$.  Thus the
canonical specific heat $c_{bulk}(T)$ will be positive and approach finite
values at the transition of first order.  At $T=T_{tr}$ $c_{bulk}(T)$ has an
additional peak $=q_{lat}\delta(T-T_{tr})$.

In the canonical ensemble the phase transition is a sharp transition (the sharp
$\delta(T-T_{tr})$ in the specific heat) only in the thermodynamic limit.
However, as a function of the specific energy the microcanonical ensemble
transforms smoothly from one phase over the coexistence region into the other
phase, see above, even in the thermodynamic limit. The nonvanishing width in
energy of the phase coexistence persists even in the thermodynamic limit.
Therefore it seems to me the singularity of $c(T)$ reflects more a ---
sometimes even unpleasant --- mathematical feature of the canonical ensemble to
suppress inhomogeneities like phase coexistence than it does illuminate the
underlying physics of the transition. This is completely contained in the
(smooth) microcanonical equation of state. In fact it is exactly here where
important information becomes hidden in the canonical ensemble.

Even though our conlusion from the Potts model is very convincing esp. in view
of our remarks to eq.(\ref{binder}), it is not a general proof that the S-bend
of the microcanonical caloric curve is neccessary for a phase transition of
first order.  S. Berry and others give some sufficient conditions for the
backbending of the caloric curve in several model systems
\cite{kunz94,wales95a,haberland94,lyndenbell95a,doye95}. These give a lot of
insight into the mechanism that produces the S-bend of $<T(E/N)>$. Berry uses
Nos\'{e}-molecular dynamics to work with intensive variables like the
temperature whereas we emphasize the advantages of the microcanonical ensemble
and the use of the conserved extensive quantities like the total energy per
particle $\varepsilon$ which are better suited especially for the study of
phase separation.

If the specific latent heat $q_{lat}\mlora 0$ and the specific interphase
surface entropy $\Delta s_{surf}\mlora 0$ the caloric equation of state gets
only a saddle point at the transition. Then $<E(T)>/N$ as well as
$<T(E/N=\varepsilon)>$ become single valued, the transition is continuous in the
canonical as well as in the microcanonical ensemble. We have a phase transition
of second order.  The two poles of $c(\varepsilon)$ merge and $c(\varepsilon)$
or $c_{bulk}(T)$ has a singularity at the transition point
$\varepsilon_{tr}$,$T_{tr}$. Both slopes of $c(\varepsilon)$ are fully
accessible in the canonical treatment of the heat capacity.
Consequently, from the microcanonical caloric equation of state
$T(\varepsilon)$ it is possible to identify and distinguish both kinds of
transitions. In microcanonical thermodynamics the close relation between the
two is very natural, transparent and simple.

It is further instructive that in finite realizations of the two-dimensional
Potts model with $q=10$ spin orientations at each lattice point it was
not possible to see a clean separation into a compact region of ordered spins
and a compact region of disordered spins at energies inside the coexistence
region even for a lattice of $100*100$ points. There were always several ``gas
bubbles'' and ``droplets'' fluctuating over the lattice and prohibiting large
interphase surfaces. A clean interphase surface, another signal of a first
order phase transition in bulk systems, cannot be seen in such small systems.
Nevertheless the caloric equation of state $T(\varepsilon)$ is already close to
its asymptotic form. (This is also the reason why it is possible to study
many-body phenomena like phase transitions in relatively small systems with
some $100$ atoms like we do here.) We can conclude from this observation that
the other classical signal of a transition of first order, {\em a clear
separation of the two phases}, is not a useful signal of a transition of first
order in small systems.
\subsection{Phase separation of macroscopic systems in the canonical
ensemble\label{phsep}}
According to eq.\ref{binder} inhomogeneous configurations like configurations
with two separated phases become exponentially suppressed by a factor
$exp(-\Delta\sigma_{surf}N^{2/3}/T)$ in the canonical ensemble. Nevertheless
such configurations are frequently treated by conventional thermodynamics with
the help of the Maxwell-construction. This is a method to construct the {\em
microcanonical, not the canonical} ensemble at energies where both phases
coexist: Homogeneous configurations with $N\nu_l$ particles in the lower phase
(``liquid'') are linked to homogeneous configurations with $N\nu_g$ particles
in the energetically higher (``gas'') phase by the condition of equal chemical
potentials $\mu_l=\mu_g$ with $\mu=T\partial \mbox{GF}/\partial N $ along a
well defined interphase surface. After taking the thermodynamic limit $N\mra
\infty$ these homogeneous configurations can be obtained from the homogeneous
canonical distributions at the lower (higher) temperature in the limit $T\mra
T_{tr}^-$ ($T\mra T_{tr}^+$). The mean total
energy can be split into the energy of the homogeneous liquid with specific
energy $\nu_l\varepsilon_l$ and the homogeneous gas part with specific energy
$\nu_g\varepsilon_g$ and vanishing fluctuations.  Consequently the fluctuation
of the total energy vanishes as well and the combined ensemble is the {\em
microcanonical} ensemble with the specific total energy
$\varepsilon=\nu_l\varepsilon_l+\nu_g\varepsilon_g$.

This is in contrast to a situation
where the total system is treated as a single canonical ensemble without a
forced phase separation. There the fluctuation of the total energy per particle
remains proportional to the specific latent heat.
In the latter case configurations with phase separation disappear.

Clearly this method works only if an external field like the gravitational
field orders the liquid below a given surface and the gas above. In a free
cluster such a simple interphase surface can usually not be fixed and the above
method to circumvent the exponential suppression of the inhomogeneous
configuration with separated phases (liquid and gas or solid and liquid) by the
canonical ensemble is not possible.
 
\section{Phase transition towards cluster fragmentation
\label{phasetransit}}
There is a lot of interest in cluster fragmentation.  Most of the work done
considered the dynamical evolution of fragmentation.  Starting with the
atom-atom interaction one studies the explicit time-evolution of some small
clusters with the help of Molecular Dynamics ($M\!D$)
\cite{rao87,barnett91,brack93,lopez94,yannouleas95,yannouleas95a,rhomund96}.
We know from other systems that the dynamical evolution of a complicated
many-body system is very much guided by the accessible phase-space. Then a
statistical treatment explains the main outcome of such reactions. Moreover, a
thermodynamic approach of cluster fragmentation offers the immediate connection
to the thermodynamics of the bulk and especially to the most interesting
thermodynamic phenomena: phase transitions. With finite clusters one can study
the development of structural transitions like melting with increasing number
of atoms see for example the illuminating review article by S.Berry in
\cite{haberland94} or the work of
\cite{doye95b,kunz94,lyndenbell94,lyndenbell95a,doye95}.

Most of these studies have been performed by embedding the cluster in a
constant temperature heat bath. This is done in $M\!D$ by using the Nos\'{e}
dynamics. From what was discussed above in the example of the Potts model the
use of the canonical ensemble to describe phase transitions in finite clusters
is not only unnatural but it seems especially inconvenient for discussing phase
coexistence. We saw that inhomogeneous configuration are suppressed $\propto
exp(-\sigma_{surf}N^{2/3}/T)$ relative to configurations with homogeneous
phases in constant temperature realizations. Presumably
this exponential suppression is the main reason for the vanishing of $\Delta
T_c=T_m-T_f$, the difference of the melting and freezing temperatures for large
systems ($N \mra \infty $) 
discussed by Berry \cite{haberland94}. Here 
microcanonical thermodynamics offers a promising and quite natural alternative
to investigate this problem.

Fragmentation is a new phase transition of finite systems. As other phase
transitions of first order it is intimitely linked to the abilitiy of finite
systems to become inhomogeneous \cite{gross153}. For infinitely many atoms it
may become the liquid-gas transition. This will be the topic of the fourth
paper in this series \cite{gross157}.  Here we are not so much interested in
the well known physics of macroscopic phase-transitions but more in phase
transitions and thermodynamics of finite systems.

The thermodynamics of isolated metal clusters has many similarities with
that of hot nuclei see ref.\cite{gross155}.  A new aspect of atomic cluster
fragmentation, however, is the fact that charge and mass degrees of freedom are
nearly decoupled. This is in sharp contrast to nuclear fragmentation where
neutrons and protons are strongly coupled by the symmetry energy.  It leads to
a predominantly symmetric nuclear fission whereas for example doubly charged
alkali clusters fission highly asymmetrically until strong shell effects of the
daughters favor a more symmetrical split. The conditions for symmetric versus
asymmetric fission will be discussed in the next paper of this series
\cite{gross151}.  A further peculiarity of clusters plays an important role in
fragmentation: 
Atomic clusters have a much higher internal density of states than nuclei. The
internal entropy $s_{int}$ can easily become larger than $10$ per atom whereas
in nuclei it is rarely $s_{int}>2$ per nucleon.  This difference is of course
due to the different statistics, fermionic statistics in a nucleus and bosonic
statistics for the phonons in the cluster.  Moreover, the internal structure of
larger clusters has significant transitions like melting at which the internal
spectrum shows strong enharmonicities.  The bulk specific heat $c_p(T)$ is
often considerably larger near the melting transition than the Debye limit for
harmonic vibrations see fig.\ref{cvNa}. As shown in
refs.\cite{gross141,gross140} this additional internal entropy leads to an
enhancement of fragmentation and also may enlarge the evaporation times by up
to three orders of magnitude.

As the charge and mass degrees of freedom are nearly decoupled in alkali
clusters, it is possible to drive the system through the phase transition not
only by increasing the excitation energy $\varepsilon$ but also by increasing
its charge $Z$. Multiply charged but cold alkali clusters may Coulomb-explode,
e.g.\cite{chandezon95,seifert96}.  This is the topic of the third paper in this
series \cite{gross152}. A third way to explore the fragmentation transition
would be to subject the cluster to large rotations and disrupt it by the
centrifugal force. All three conserved quantities, the energy $E$, the charge
$Z$, and the angular momentum $\vecb{L}$ are possible nonfluctuating order
parameters of this transition. At present it is experimentally not possible to
transfer a large known angular momentum to a cluster. In the analogous case of
nuclear fragmentation the disruptive effect of large rotations has just started
to be discussed see e.g. \cite{gross143}. In a couple of papers Jellinek and
collaborators explored very early the effect of rotation on the isomerization
and the behavior of clusters at low temperatures
\cite{li88a,jellinek89,jellinek90,lopez94}. Their interest aims into a similar
direction as ours. They are concerned with the important problem how the low
energetic properties of a cluster becomes modified by the rotation. We here
want to study the effect of the two-dimensional stress by the centrifugal force
on the fragmentation. It is further important to realize that the longrange
centrifugal force induces inhomogeneities in the system which again demand a
statistical treatment by microcanonical thermodynamics which leads to
interesting pecularities as a kinetic temperature that can considerably exceed
the thermodynamic temperature \cite{gross143}.

We sample the accessible N-body phase space by {\em M}icrocanonical 
{\em M}etroplis {\em M}onte {\em C}arlo, $M\!M\!M\!C$. Its basic idea is as
follows: Since every quantum state of the N-body system has an equal
probability in the microcanonical ensemble one has to sample each quantum
state. There is no hierarchy of a few important ones representing the main
behavior of the system.  In view of the tremendously large number of
participating states this is impossible to do.  Therefore it is essential to
group the states by some common quality. In most experiments there are several
observables which remain undetermined. In a typical atomic cluster experiment
one counts the number of charged fragments but usually does not specify their
internal excited state. Also one will usually not determine the momentum of
each neutral evaporated monomer.  Consequently, the natural sorting principle
is the size of the unobserved phase space. As explained in great detail in
refs.\cite{gross141,gross95,gross153} $M\!M\!M\!C$ samples only the relevant
number of degrees of freedom. It treats the internal degrees of freedom of the
fragments using the known internal specific heat. Each set of explicit degrees
of freedom represents a whole subset of quantum states of the system and is
weighted by the size of the corresponding part of the phase space. There are
only a few millions of different configurations which are important for a
specific reaction.  $M\!M\!M\!C$ gives a systematic method to find them. It is
thus possible to determine the size of the accessible N-body phase space i.e.
the total entropy which is rather difficult to get e.g. by molecular dynamics.

Let us now consider the fragmentation of sodium clusters with
increasing number of atoms and charges. Figure \ref{na2-100} shows the
average mass of the three largest fragments versus the specific excitation
energy for the fragmentation of Na$^{2+}_{100}$. The thick solid curve gives
the thermodynamic temperature in Kelvin (right scale). Below $\varepsilon =0.5$
eV/atom a singly charged big ``evaporation'' residue with dominating mass and
additionally a small charged as well as several small neutral fragments are
produced. Above $0.5$eV/atom we find two small singly charged and several
uncharged fragments of similar size ($\le 10$) (multifragmentation).  The
caloric equation of state $<T(E/N)>$ shows a significant backbending, which is
a clear signal for a phase transition of first order at
$\varepsilon\approx 0.5$eV/atom. The two phases are characterized by the
presence or absence of a dominating large fragment.

Doubling the size of the cluster gives an even more pronounced phase
transition, fig.\ref{na2-200}. The transition is shifted to slightly higher
excitation energies.  Evidently, fragmentation is easier in a
smaller cluster with a large surface to volume ratio. The behavior of the mass
distribution with excitation energy is very similar to the smaller cluster.
In fig.\ref{na2-500} we show the fragmentation of an even larger cluster,
Na$^{2+}_{500}$. The mass distribution is like in the two previous cases, the
transition is shifted further up to $\approx 0.7$eV/atom.

Rising the charge from $Z=2+$ to $Z=5+$ shifts the transition point again down
to $\varepsilon=0.55$ eV/atom, fig.\ref{na5-200}. Rising the charge has a
similar effect on the fragmentation transition as dropping the size of the
cluster. All calculations were performed in the conducting sphere approximation
to include the polarization of the charge distribution of Na$_n$, see paper
III \cite{gross152} of this series for a detailed explanation of the method.

Figure \ref{na5pc200} shows a calculation using only the monopole-monopole
(point charge) Coulomb interactions without the higher multipoles by the
induced image charges. The transition is shifted towards slightly higher
energies. The polarization (mirror charges) induces a net attraction. In
contrast to similar calculations in ref.\cite{gross141} we used here the
experimental binding energies for Na$^{q}_n$ with $q=0,1+$ and $2\le n\le 21$
given by ref.\cite{kappes88,brechignac89} , whereas in \cite{gross141} we had
only experimental values for neutral and charged dimers and trimers. This
change induces a stronger upwards shift of the transition than the difference
between conducting and point-charge distribution. The simple reason is the
predominant production of small fragments with masses between $n=1$ and $n=10$,
so that the transition is especially sensitive to small changes in the binding
energies of these small fragments.
\section{Conclusion}
Microcanonical thermodynamics is the basics of all thermodynamics, but only
recently we came into the position to explore it for nontrivial situations. In
contrast to the more familiar canonical thermodynamics it is the only
formulation which is valid for small isolated systems like free atomic
clusters. We can now study how phase transitions evolve in systems with rising
number of atoms from small towards the macroscopic ones.

Fragmentation phase transition is a new but generic phase transition of first
order in small systems. Like other more familiar first order transitions as
melting or boiling they lead to strong spatial inhomogeneities. However,
unlike to the macroscopic transitions as boiling under the presence of an
external gravity field, the inhomogeneities in microcanonical small systems are
strongly fluctuating and the partition entropy plays an important role. Besides
of being the correct ensemble for isolated systems the microcanonical ensemble
is more suited to describe configurations with eventual phase separation than
the canonical ensemble in which such configuration are exponentially suppressed
$\propto exp(-N^{2/3}\sigma_{surf}/T)$. A fact that is especially important
if one wants to study the evolution with rising particle numbers of
configurations with two coexistent phase as e.g. the solid clusters with a
molten surface or for the evolution of the liquid -- gas transition as will be
discussed in \cite{gross157}.

The fragmentation transition was first described for the multifragmentation of
hot nuclei \cite{gross45,bondorf81,gross95,gross155}.  In this series of papers
we will discuss the fragmentation phase transition in clusters under various
external conditions. As we have seen this transition is well defined in sodium
clusters of some 100 atoms. With rising size and increasing volume to surface
ratio it shifts towards higher excitation per atom.  Charging the cluster leads
to a lowering of the critical excitation energy per atom.

In contrast to nuclear multifragmentation, the fragmentation of alkali clusters
is very asymmetric if not special magic shell closures in the fragments favor
more symmetric splitting. This topic will be discussed in the next paper of
this series \cite{gross151}.

All our results concern the topological structure of the N-atom phase space
only. Whether a realistic dynamical system can explore all the topological
details of the phase space depends on how ergodic the dynamics is. The strong
friction between nuclei moving one against another at close distances certainly
helps to equilibrate the various degrees of freedom in nuclear fragmentation.
Whether there is a similar strong friction between moving atomic clusters is
not known.  However, this is to be expected since atomic clusters have normally
a less compact surface than nuclei.

The basic limitations and approximations which we had to use in our analysis
are discussed in \cite{gross155} and will be listed at relevant places in the
following papers.

\section{Acknowledgments} 
D.H.E.G. is very gratefull to A. H\"uller, S. Gro\ss mann, S. Berry and J.
Jellinek for enlightning discussions about the basic principles of
thermodynamics.  We thank the Sonderforschungsbereich SFB 337 of the Deutsche
Forschungsgemeinschaft for substantial support. We are also grateful to the
Freie Universit\"at Berlin for supporting us with computer time.
%\bibliographystyle{unsrt}%{alpha}%{alpha}%{plain} %{unsrt}
%\bibliography{c:/bibliogr/gross,c:/bibliogr/othbibam,c:/bibliogr/othbibnz}

\clearpage
\begin{thebibliography}{10}

\bibitem{ellert95}
C.~Ellert, M.~Schmidt, C.~Schmitt, T.~Reiners, and H.~Haberland.
\newblock Temperature dependence of the optical response of small open shell
  sodium clusters.
\newblock {\em Phys. Rev. Lett}, 75:1731, 1995.

\bibitem{gross157}
D.H.E. Gross and M.E. Madjet.
\newblock Fragmentation phase transition in atomic clusters IV --- the relation
  of the fragmentation phase transition to the bulk liquid-gas transition.
\newblock {\em HMI-preprint}, in preparation, 1996.

\bibitem{thompson88}
C.J. Thompson.
\newblock {\em Classical Equilibrium Statistical Mechanics}.
\newblock Clarendon Press, Oxford, 1988.

\bibitem{boltzmann}
Ludwig Boltzmann.
\newblock {\em Vorlesung \"uber Gastheorie}.
\newblock Number~1. Akademische Druck-u. Verlagsanstalt, Graz, 1981.

\bibitem{jellinek86}
J.~Jellinek, T.L. Beck, and R.S. Berry.
\newblock Solid-liquid phase changes in simulated isoenergetic ar$_{13}$.
\newblock {\em J.Chem.Phys.}, 84:2783, 1986.

\bibitem{vanhove49}
L.~van Hove.
\newblock Quelques propri$\acute{e}$t$\acute{e}$s g$\acute{e}$n$\acute{e}$rales
  de l'int$\acute{e}$grale de configuration d\'\,un syst$\grave{e}$me de
  particules avec interaction.
\newblock {\em Physica}, 15:951, 1949.

\bibitem{baxter73}
R.J. Baxter.
\newblock {\em J. Phys.}, C6:L445, 1973.

\bibitem{gross150}
D.H.E. Gross, A.~Ecker, and X.Z. Zhang.
\newblock Microcanonical thermodynamics of first order phase transitions
  studied in the potts model.
\newblock {\em Ann. Physik}, 5:446--452, 1996.

\bibitem{hueller94}
A.~H\"uller.
\newblock Finite size scaling at first order phase transitions ?
\newblock {\em Z.Phys.B}, 95:63--66, 1994.

\bibitem{promberger96}
M.~Promberger.
\newblock On a trivial aspect of canonical specific heat scaling.
\newblock {\em preprint, Erlangen}, 1996.

\bibitem{fisher67}
M.E. Fisher.
\newblock {\em Physics}, 3:255, 1967.

\bibitem{doye95b}
J.P.K. Doye, D.J. Wales, and R.S. Berry.
\newblock The effect of the range of the potential on the structures of
  clusters.
\newblock {\em J. Chem. Phys.}, 103:4234--4249, 1995.

\bibitem{kunz94}
R.~E. Kunz and R.S. Berry.
\newblock Multiple phase coexistence in finite systems.
\newblock {\em Phys.Rev.}, E49:1895, 1994.

\bibitem{wales95a}
D.J. Wales and J.P.K. Doye.
\newblock Coexistence and phase separation in clusters: From the small to the
  not-so-small regime.
\newblock {\em J. Chem. Phys.}, 103:3061--3070, 1995.

\bibitem{binder82}
K.~Binder.
\newblock Monte carlo calculation of the surface tension for two- and
  three-dimensional lattice-gas models.
\newblock {\em Phys. Rev. A}, 25:1699--1709, 1982.

\bibitem{haberland94}
H.~Haberland (Ed).
\newblock {\em Clusters of Atoms and Molecules}.
\newblock Springer-Verlag, Berlin, Heidelberg, New York, 1994.

\bibitem{lyndenbell95a}
R.M. Lynden-Bell.
\newblock Negative specific heat in clusters of atoms.
\newblock {\em to be published in Galactic Dynamics}, 1995.

\bibitem{doye95}
J.P.K. Doye and D.J. Wales.
\newblock An order parameter aproach to coexistence in atomic clusters.
\newblock {\em J. Chem. Phys.}, 102:9673--9688, 1995.

\bibitem{rao87}
B.K. Rao, P.~Jena, M.~Manninen, and R.M. Nieminen.
\newblock Spontaneous fragmentation of multiply charged metal clusters.
\newblock {\em Phys. Rev. Lett.}, 58:1188--1191, 1987.

\bibitem{barnett91}
R.N. Barnett, U.~Landman, and G.~Rajagopal.
\newblock Patterns and barriers for fission of charged small metal clusters.
\newblock {\em Phys.Rev.Lett.}, 67:3058, 1991.

\bibitem{brack93}
M.~Brack.
\newblock {\em Rev.Mod.Phys.}, 65:677, 1993.

\bibitem{lopez94}
M.J. L\'opez and J.~Jellinek.
\newblock Fragmentation of atomic clusters: A theoretical study.
\newblock {\em Phys.Rev.A}, 50:1445, 1994.

\bibitem{yannouleas95}
C.~Yannouleas and Uzi Landman.
\newblock {\em Phys. Rev. B}, 51:1962, 1995.

\bibitem{yannouleas95a}
C.~Yannouleas, R.N. Barnett, and Uzi Landman.
\newblock Electronic shell effects in fission barriers and fission dynamics of
  metal clusters.
\newblock {\em Com.At.Mol.Phys.}, 31:445--460, 1995.

\bibitem{rhomund96}
F.~Rhomund, E.E.B. Campbell, O.~Knospe, G.~Seifert, and R.~Schmidt.
\newblock Collision energy dependence of molecular fusion and fragmentation in
  C$_{60}^+$ + C$_{60}$ collisions.
\newblock {\em Phys. Rev. Lett.}, in print, 1996.

\bibitem{lyndenbell94}
R.M. Lynden-Bell and D.J. Wales.
\newblock Free energy barriers to melting in atomic clusters.
\newblock {\em J. Chem. Phys.}, 101:1460--1476, 1994.

\bibitem{gross155}
D.H.E. Gross.
\newblock Microcanonical thermodynamics, fragmentation ``phase-transition'',
  and the topology of the n-body phase space.
\newblock In S.Albergio, S.Costa, A.Insolia, and C.Tuve, editors, {\em
  Proceedings of CRIS96 ''Critical Phenomena and Collective Observables''},
 http://xxx.lanl.gov/nucl--th/9607038,  
 Acicastello, Sicily, Italia, 27.5.-31.5.96, 1996. World Scientific,
  Singapore.

\bibitem{gross151}
M.E. Madjet, D.H.E. Gross, P.A. Hervieux, and O.~Schapiro.
\newblock Fragmentation phase transition in atomic clusters II --- symmetry of
  coulombic fission.
\newblock {\em HMI-preprint}, in preparation, 1996.

\bibitem{gross141}
D.H.E. Gross and P.A. Hervieux.
\newblock Statistical fragmentation of hot atomic metal clusters.
\newblock {\em Z. Phys. D}, 35:27--42, 1995.

\bibitem{gross140}
P.A. Hervieux and D.H.E. Gross.
\newblock Evaporation of hot mesoscopic metal cluster.
\newblock {\em Z. Phys.D}, 33:295--299, 1995.

\bibitem{chandezon95}
F.~Chandezon, C.~Guet, B.A. Huber, D.~Jalabert, M.~Maurel, E.~Monnand,
  C.~Ristori, and J.C. Rocco.
\newblock {\em Phys. Rev. Lett.}, 74:3784, 1995.

\bibitem{seifert96}
G.~Seifert, R.~Gutierrez, and R.~Schmidt.
\newblock Ionization energies and coulomb explosion of highly charged c$_{60}$.
\newblock {\em Phys. Lett.}, A 211:357, 1996.

\bibitem{gross152}
O.~Schapiro, P.J. Kuntz, K.~M\"ohring, P.A. Hervieux, M.E. Madjet, and D.H.E.
  Gross.
\newblock Fragmentation phase transition in atomic clusters III --- coulomb
  explosion of cold clusters.
\newblock {\em HMI-preprint}, in preparation, 1996.

\bibitem{gross143}
A.S. Botvina and D.H.E. Gross.
\newblock The effect of large angular momenta on multifragmentation of hot
  nuclei.
\newblock {\em Nucl.Phys.}, A 592:257--270, 1995.

\bibitem{li88a}
D.H. Li and J.~Jellinek.
\newblock Rotating clusters: centrifugal distortion, isomerization,
  fragmentation.
\newblock {\em Z.Phys.D}, 12:177--180, 1988.

\bibitem{jellinek89}
J.~Jellinek and D.H. Li.
\newblock Separation of the energy of overall rotation in any n-body system.
\newblock {\em Phys.Rev.Lett}, 62:241, 1989.

\bibitem{jellinek90}
J.~Jellinek and D.H. Li.
\newblock Vibrations of rapidly rotating n-body systems.
\newblock {\em Chem.Phys.Lett.}, 169:380, 1990.

\bibitem{gross95}
D.H.E. Gross.
\newblock Statistical decay of very hot nuclei, the production of large
  clusters.
\newblock {\em Rep.Progr.Phys.}, 53:605--658, 1990.

\bibitem{gross153}
D.H.E. Gross.
\newblock Microcanonical thermodynamics and statistical fragmentation of
  dissipative systems --- the topological structure of the n-body phase space.
\newblock {\em Physics Reports}, in preparation, 1996.

\bibitem{kappes88}
M.M. Kappes, M.~Sch\"ar, U.~R\"othlisberger, G.~Yeretzian, and E.~Schumacher.
\newblock Sodium cluster ionisation potentials revisited: Higher-resolution
  measurements for na$_n$ ($n<23$) and their relation to bonding models.
\newblock {\em Chem. Phys. Lett.}, 143:251, 1988.

\bibitem{brechignac89}
C.~Br\'echignac, Ph. Cahuzac, J.~Leygnier, and J.~Weiner.
\newblock Dynamics of unimolecular dissociation of sodium cluster ions.
\newblock {\em J. Chem. Phys.}, 90:1492, 1989.

\bibitem{gross45}
D.H.E. Gross and Meng Ta-chung.
\newblock Production mechanism of large fragments in high energy nuclear
  reactions.
\newblock In {\em 4th Nordic Meeting on Intermediate and High Energy Physics},
  page~29, Geilo Sportell, Norway, January 1981.

\bibitem{bondorf81}
J.P. Bondorf.
\newblock A model for fragmentation in intermediate energy heavy ion reactions.
\newblock In C.H. Dasso, editor, {\em Nuclear Physics, Proceedings of the
  Nuclear Physics Workshop, I.C.T.P., Trieste, Miramare, Italy, 5-30 October,
  1981}, pages 765--770, Amsterdam, New York, Oxford, 1982. North Holland.

\bibitem{borelius63}
G.~Borelius.
\newblock The changes in energy content, volume, and resistivity with
  temperature in simple solids and liquids.
\newblock {\em Solid State Physics}, 15:1, 1963.

\bibitem{hultgren63}
R.~Hultgren, R.L. Orr, P.D. Anderson, and K.K. Kelley.
\newblock {\em Selected Values of Thermodynamic Properties of Metals and
  Alloys}.
\newblock John Wiley ans Sons, New York, 1963.

\end{thebibliography}
\clearpage
\begin{figure}
\caption{Specific heat of bulk sodium at atmospheric pressure from  
\protect\cite{borelius63,hultgren63}. The dashed line represents the specific
heat calculated within the Debye model. $T_m$ is the melting and $T_v$ is the 
boiling transition. Notice the overshooting of $c(T)$ over the Debye limit of
harmonic vibrations at the melting point.}
\label{cvNa}
\end{figure}
\begin{figure}
\caption{a) Specific entropy
\protect{$s(\varepsilon)=\int_0^{\varepsilon}{\beta_{micro}(\bar{\varepsilon})
d\bar{\varepsilon}}$} versus  the specific energy $\varepsilon$ for the 2-dim.
Potts model with $q=10$ on a $100*100$ lattice.  In order to visualize the
anomaly of the entropy the linear function $a+b\varepsilon$ ($a=0.2119$,
$b=1.4185$) was subtracted. Because we use periodic boundary conditions one
needs two cuts to separate the phases  and the depth of the convex intruder is
twice the surface-entropy.
\protect\newline
b) Microcanonical caloric equation of state,
$\beta_{micro}(\varepsilon)=1/T(\varepsilon)$ as directly calculated by
$M\!M\!M\!C$
\protect\newline
c) Specific heat $c(\varepsilon)=-\beta^2/(\partial\beta/\partial\varepsilon)$.
The canonical ensemble of the bulk jumps over the shaded region between the
vertical lines at $\varepsilon_1$ and $\varepsilon_3$. This is the region of
the coexistence of two phases one with ordered spins, the other with disordered
spins. Here $c(\varepsilon)$ has two poles and becomes negative in-between.
The canonical thermodynamics is blind to this region. Notice that the poles
are {\em inside} $\varepsilon_1\le\varepsilon\le\varepsilon_3$, i.e the
canonical specific heat remains finite and positive as it should.}
\label{entrop} 
\end{figure}
\begin{figure}
\caption{Average masses of the three largest fragments and the thermodynamic
temperature T as function of the excitation energy per atom for the
fragmentation of Na$^{2+}_{100}$ .\label{na2-100} }
\end{figure}
\begin{figure}
\caption{Same as figure \protect\ref{na2-100} but for
Na$^{2+}_{200}$.\label{na2-200} }
\end{figure}
\begin{figure}
\caption{Same as figure \protect\ref{na2-100} but for
Na$^{2+}_{500}$.\label{na2-500} }
\end{figure}
\begin{figure}
\caption{Same as figure \protect\ref{na2-100} but for
Na$^{5+}_{200}$.\label{na5-200} }
\end{figure}
\begin{figure}
\caption{Same as figure \protect\ref{na2-100} but for
Na$^{5+}_{200}$ and point charge Coulomb interaction.\label{na5pc200} }
\end{figure}
%\end{document}

\end{document}